\documentclass[12pt]{iopart}
\usepackage[usenames]{color}
\usepackage{ae,aecompl}
\usepackage[normalem]{ulem}
\usepackage{graphicx}
\usepackage[utf8]{inputenc}
\begin{document}

\title[Microinstability destabilization from shielded 3-D magnetic perturbations]{A model for microinstability destabilization and enhanced transport in the presence of shielded 3-D magnetic perturbations}

\author{T M Bird$^1$, C C Hegna$^2$}
\address{$^1$ Max-Planck-Institut für Plasmaphysik, D-17491 Greifswald, Germany}
\address{$^2$ Departments of Engineering Physics and Physics,  \\
University of Wisconsin, Madison, WI 53706}
\ead{tbird@ipp.mpg.de}

\begin{abstract}
A mechanism is presented that suggests shielded 3-D magnetic perturbations can destabilize
microinstabilities and enhance the associated anomalous transport.   Using local 3-D equilibrium theory, shaped tokamak equilibria with small 3-D deformations are constructed.  In the vicinity of rational magnetic surfaces, the infinite-n ideal MHD ballooning stability boundary is strongly perturbed by the 3-D modulations of the local magnetic shear associated with the presence of near-resonant Pfirsch-Schluter currents.  These currents are driven by 3-D components of the magnetic field spectrum even when there is no resonant radial component.   The infinite-n ideal ballooning stability boundary is often used as a proxy for the onset of virulent kinetic ballooning modes (KBM) and associated stiff transport.  These results suggest that the achievable pressure gradient may be lowered in the vicinity of low order rational surfaces when 3-D magnetic perturbations are applied.  This mechanism may provide an explanation for the observed reduction in the peak pressure gradient at the top of the edge pedestal during experiments where edge localized modes have been completely suppressed by applied 3-D magnetic fields.

\end{abstract}
\maketitle

\section{Introduction}

The use of applied three-dimensional magnetic fields can have beneficial effects on the performance of otherwise axisymmetric toroidal confinement devices \cite{boozer,callen}. In particular, the application of external 3-D resonant magnetic perturbations (RMP) whose resonant surface lies in the edge region of H-mode tokamaks can suppress the appearance of edge localized modes (ELMs) under certain conditions \cite{evans1,evans2,evans3}.   ELMs repeatedly and violently eject hot plasma from the tokamak edge onto solid materials of the device.  In order to safely operate at the desired parameters without prohibitive erosion of first wall materials, future experiments such as ITER will require a reliable method for controlling edge properties to prevent large type-I ELMs  \cite{loarte}. The original motivation for pursuing the RMP method of ELM suppression was based on the expectation that resonant magnetic perturbations of sufficient magnitude would produce overlapping magnetic islands that would greatly enhance plasma transport in the edge region and accordingly reduce the drive for ELM inducing instabilities associated with edge pressure gradients and currents \cite{snyder,wilson}.  However, plasma flows in the edge region may be of sufficient magnitude to shield resonant magnetic perturbations from penetration \cite{fitzpatrick,marsf_shielding} in which case stochastic transport is no longer viable.   Hence, a different mechanism may be required to explain the observations which indicate that the externally applied 3-D perturbations can enhance plasma transport in the pedestal and preclude the triggering of ELMs.  

Experiments with applied 3-D magnetic perturbations have now been performed on a number of tokamaks worldwide, yielding varying results.  Experiments performed at the DIII-D tokamak with $n=3$ perturbations have demonstrated suppression of large type-I ELMs at various collisionalities and with different plasma shaping.  In many cases the large type-I ELMs are replaced by higher frequency, lower amplitude edge transport bursts \cite{moyers}.  This has also been achieved for example at JET \cite{Liang} with $n=1$ and $n=2$ perturbations, at MAST with $n=3$ perburations \cite{denner}, and at ASDEX-U with $n=2$ perburbations \cite{suttrop}.  However, in some DIII-D experiments a complete elimination of bursty edge transport events has even been achieved \cite{evans3}.  In this work we provide a possible explanation for ELM-free operation that does not rely on stochastic magnetic fields to enhance the transport in the edge region of H-mode tokamaks.

We propose a model for the enhanced transport in the edge region that is based on the three-dimensional distortions of the magnetic surfaces adversely affecting microinstabilities.   While the effects of 3-D fields on stability discussed here is generic to a broad range of localized modes, we concentrate on the properties of kinetic ballooning modes (KBMs) in this work,  which are primary candidates for explaining edge pedestal transport in H-mode plasmas.  KBM growth rates become very large as the infinite-n ideal ballooning boundary is approached \cite{kbm,pueschel,belli}.  As such, ideal MHD ballooning mode calculations are often used as a proxy for the abrupt onset of KBM instabilities and the associated stiff transport response \cite{EPED}. In particular, calculations using  information from ideal ballooning  stability and  peeling-ballooning (P-B) theory used to predict ELM onset have successfully predicted self-consistent pedestal widths and heights \cite{groebner}.  Gyrokinetic modeling of microinstabilities in the pedestal also supports the hypothesis that KBMs limit the pedestal pressure gradient and thus regulate the inward advance of the pedestal \cite{dickinson1,dickinson2}.

In this work, the effect of the 3-D distortions of the MHD equilibrium on ideal ballooning stability is calculated.   For amplitudes of comparable measure to those present in experiments, the 3-D distortions are shown to substantially modify the local shear and reduce the critical pressure gradient for ideal ballooning instability.    This is purely a 3-D MHD equilibrium effect, occuring due to Pfirsch-Schlüter currents near rational surfaces that are driven by 3-D inhomogeneity in
the magnetic field spectrum. This is a resonant effect and would lower the achievable pressure gradient in the vicinity of low order rational surfaces where the 3-D deformation is strongest.  Enhanced transport due to the onset of KBM instabilities at lower pressure gradient, due to the 3-D effect, could halt the inward progression of the edge pedestal and achieve ELM mitigation by preventing the pedestal height from reaching values where P-B modes are driven unstable.  This explanation is consistent with recent pedestal modeling work where the prevention of ELMs was explained by the existence of a 'wall' at a low order rational surface which prevents the inward advance of the pedestal \cite{snyder2}.  

It is worth noting that the global MHD instabilities that are thought to trigger ELMs, Peeling-Ballooning modes, may be sensitive to this local shear modulation as well.  The high-n components of these modes are dominated by ballooning structure, and we show in this work that (local) infinite-n ballooning modes are strongly affected by the local shear modulation (though we use the infinite-n ballooning calculations merely as a proxy for the onset of KBM instabilities and the associated transport).  To understand the effect of resonant Pfirsch-Schlüter currents on the Peeling-Ballooning instabilities would require global MHD stability calculations using global 3-D MHD equilibria which include the resonant Pfirsch-Schlüter physics.  This is beyond the scope of this work.  In particular, this model attempts to provide an explanation for the DIII-D experiments where the increased pedestal transport completely precludes the triggering of any ELMs or other bursty events.  Therefore, we focus on local KBM stability (with infinite-n MHD ballooning mode stability as a proxy) and utilize radially local equilibrium, which greatly simplifies the modeling effort and should be adequate for microinstability calculations.  

In the following section, details for how 3-D fields alter the MHD equilibrium are
described using local 3-D equilibrium theory.  In section III, the infinite-n ideal MHD ballooning stability boundary is evaluated for a set of equilibria with 3-D flux surface deformations of experimentally relevant magnitude.  It is found that as the q value approaches a rational value, the instability boundary is strongly modified by the presence of the 3-D fields.  Section IV explains this result via a detailed examination of the local magnetic shear.  The stability boundary is expanded when near-resonant Pfirsch-Schlüter currents modulate the local magnetic shear in a manner strongly conducive to ballooning instability.  Section V summarizes these results and discusses the implications for understanding experiments where RMP suppression of Edge Localized Modes has been achieved.  Appendix A elucidates the role of the 3-D fields on the local shear and provides an analytic estimate for the critical pressure gradient for ballooning instability for a low-$\beta$, high aspect ratio, circular cross section tokamak when weak 3-D fields are present.

\section{The local 3-D equilibrium model}

A description of the equilibrium in the presence of 3-D fields can be provided by the prescriptions of local 3-D MHD equilibrium theory \cite{local3D}.  In this formulation, the shape of the 3-D magnetic surface is parametrized by straight-field line angles, the value of the safety factor $q$ and two profile quantities, typically the presure gradient and the average magnetic shear $q^{\prime}$. These parameters uniquely specify the properties of 3-D MHD equilibrium in the vicinity of a flux surface. The original motivation for this formulation was to study the effects of 3-D shaping on ballooning stability in stellarator configurations \cite{hh}, and it has also proved useful for interpreting LHD data \cite{nakajima06}.  This formulation provides an exact description of how 3-D plasma shaping affects quantities such as the local magnetic shear and normal curvature which play crucial roles in the stability of localized plasma instabilities.  In this work, this model is used to elucidate the mechanism by which small 3-D deformations can introduce substantial 3-D structure in the Pfirsch-Schlüter current spectrum and local magnetic shear. 

A shaped tokamak equilibrium model, the widely used ''Miller equilibrium'' \cite{miller}, provides the starting point for these calculations.  The physical position of the flux surface in cylindrical coordinates is parametrized in terms of the geometric poloidal angle $\theta$ as
\begin{eqnarray}
& R = R_{0} + r \cos [ \theta + (\sin^{-1} \delta ) \sin \theta ] , \label{eq:Reqn}\\ 
& Z = \kappa r \sin \theta \label{eq:Zeqn}
\end{eqnarray}
where $\delta$ is the plasma triangularity, $\kappa$ is the
elongation, and $A = R_{0} / r $ is the aspect ratio of the flux
surface.  The poloidal magnetic field on the surface is also specified using four
additional parameters,
\begin{equation}
B_{p} =
 \frac{ d_{r} \psi [ \sin^{2} ( \theta + x \sin \theta ) (
  1 + x \cos \theta )^{2} + \kappa^{2} \cos^{2} \theta ]^{1/2} }
{\kappa R a_{0}} , 
\end{equation}
where $ a_{0} = \cos ( x \sin \theta ) + d_{r} R_{0} \cos \theta + [ s_{\kappa}-s_{\delta} \cos \theta+( 1 + s_{\kappa} ) x \cos \theta ] \sin
  \theta \sin ( \theta + x \sin \theta ) $, $ \sin x = \delta $ and three of the new parameters $s_{\kappa},
s_{\delta},$ and 
$d_{r} R_{0}$ are related to radial derivatives of the flux surface shaping parameters \cite{miller}. Typically q is chosen and $ d_{r} \psi $ is determined consistent with the definition $ q = ( f / 2 \pi )\int  dl_{p} / ( R^{2} B_{p} ) $ , where $ f(\psi) = R B_{\phi} $, $ dl_{p} $ is the differential poloidal arc length, and $B_{\phi}$ is the toroidal magnetic field.  The equilibrium parametrization is completed by choosing the following two dimensionless profile quantities,
\begin{eqnarray}
& s = \frac{q^{\prime} R_{0}^{3} }{q \hat{V}^{\prime} c_{0} }\\
& \alpha = - \frac{ 2 q^{2} \mu_{o} p^{\prime} \hat{V}^{\prime} R_{0} }{c_{0}^{1/2}}
\end{eqnarray}
where $\hat{V}' = \int d\Theta \int d\zeta \sqrt{g}/4\pi^2$, $\sqrt{g}$ is the Jacobian and $c_{0}=<B^{2} R_{0}^{2} / | \nabla \psi |^{2}>$ where $<Q>=
\int \int d \theta d \zeta Q \sqrt{g} / \int \int d \Theta d \zeta \sqrt{g}$.  In total there are nine dimensionless parameters which uniquely specify a solution to the MHD equilibrium equations in the vicinity of the chosen flux surface.  

While the Miller equilibrium is formulated by denoting the inverse mapping $\mathbf{X}=\mathbf{X} ( \theta ) $ and the poloidal field strength,  the 3-D local equilibrium model is formulated in terms of straight field line coordinates $ \Theta $ and
$\zeta$, such that $ d \zeta / d \Theta = q ( \psi ) $.  These two treatments can be unified by noting the relationship
\begin{equation}
B_{p} = \frac{1}{q \sqrt{g}} \left | \frac{ \partial \mathbf{X} }{\partial
  \theta} \right | \frac{ \partial \theta } { \partial \Theta } .
\end{equation}
Here, the Jacobian, $\sqrt{g} = R^{2} / f$, is written consistent with the use of symmetry in $ \zeta $.  A transformation to the straight field line coordinates used by the 3-D local model can be calculated from
\begin{equation}
\frac{ \partial \Theta}{\partial \theta} = \left | \frac{\partial
  \mathbf{X}}{\partial \theta} \right | \frac{ f }{ q R^{2} B_{p} } .
\end{equation}
 
The 3-D distortions of the flux surface shape can then be added perturbatively to the Miller equilibrium.  In the following, the flux surface parametrization given by
\begin{eqnarray}
& R = R ( \Theta ) + \sum_{i} \gamma_{i} \cos ( M_{i} \Theta - N_{i} \zeta ) , \nonumber\\
& Z = Z ( \Theta ) + \sum_{i} \gamma_{i} \sin ( M_{i} \Theta - N_{i} \zeta ) , \label{eq:param} 
\end{eqnarray}
and $ \phi = - \zeta $ will be used throughout the calculation, in which the sum is over 3-D perturbations with different helicity.  Here, $R(\Theta)$ and $Z(\Theta)$ are given by Eqns (\ref{eq:Reqn}) and (\ref{eq:Zeqn}) and the amplitudes $\gamma_{i}$ are small quantities.  

The magnitude of each $ \gamma_{i} $ in the flux surface shape parametrization can be related to the magnitude of a radial magnetic perturbation.  The resulting magnetic field is split into two components, $ \mathbf{B} = \mathbf{B}^{axi} + \mathbf{\tilde{B}} $, where $\mathbf{B}^{axi} $ is the axisymmetric part.  The perturbed part of the magnetic field is then projected into a radial coordinate, $ \rho^{2} \sim
\psi^{axi} $.  In a high aspect ratio circular cross section limit, the Fourier harmonic of each magnetic field component is given by
\begin{equation}\label{eq:Br}
\frac{\tilde{B_{\rho}}}{B_{0}}_{\left ( M = M_{i} - 1, N = N_{i}
  \right )} \cong [
q N_{i} - M_{i} ] \frac{ \gamma_{i} }{R_{0}} .
\end{equation}
Note that at a rational surface $ q = M / N $, the resonant component of the radial magnetic field is completely shielded.

Using Eq. (\ref{eq:param}) as the flux surface parametrization, the geometric properties of the magnetic field lines (normal and geodesic curvature, normal torsion, etc.) are determined.  The magnetic field and gradient in flux are given by
\begin{eqnarray}
\mathbf{B} & = \frac{1}{\sqrt{g}} \left ( \frac{ \partial
    \mathbf{X}}{\partial \zeta} + \frac{1}{q} \frac{ \partial
    \mathbf{X}}{\partial \Theta} \right ) \\
 \nabla \psi  &= \frac{1}{\sqrt{g}} \left ( \frac{ \partial
    \mathbf{X}}{\partial \Theta} \times \frac{\partial
    \mathbf{X}}{\partial \zeta} \right )
\end{eqnarray} 
where the Jacobian is defined as
\begin{equation}\label{eq:jacobian_definition}
\sqrt{g} = \frac{ \partial \mathbf{X}}{\partial \psi} \cdot \left (
  \frac{ \partial \mathbf{X}}{\partial \Theta} \times \frac{\partial
    \mathbf{X}}{\partial \zeta } \right ) .
\end{equation}
The local 3-D equilibrium model is based on an expansion in $\psi$ of the field line mapping,
\begin{equation}
\mathbf{X} ( \psi, \Theta, \zeta ) = \mathbf{X} ( \psi_{0}, \Theta,
\zeta ) + ( \psi - \psi_{0} ) \frac{ \partial \mathbf{X} }{\partial \psi
} ( \psi_{0}, \Theta, \zeta ) + ...
\end{equation}\label{eq:X}
where $\psi_{0}$ labels the surface of interest and higher order terms in the expansion are neglected.  By solving the ideal MHD equilibrium equations in the vicinity of $\psi_{0}$, $ \partial \mathbf{X}
/ \partial \psi $ can be determined.  To do this we note that the equations of MHD equilibrium dictate that no plasma currents flow normal to flux surfaces.  By calculating $ \mathbf{\hat{n}} \cdot \mathbf{J} = 0 $, where $ \mathbf{\hat{n}} $ is the unit vector normal to the flux surface, a first order partial differential equation for the Jacobian is derived,
\begin{equation}\label{eq:jacobian_eqn}
\frac{\partial}{\partial \theta} \frac{ g_{\zeta \zeta} + 
  g_{\zeta \Theta} / q}{\sqrt{g}} = \frac{\partial}{\partial \zeta} \frac{
  g_{\Theta \zeta} + g_{\Theta \Theta} / q}{\sqrt{g}} ,
\end{equation}
where the metric elements are defined as $ g_{\Theta \Theta} =
\frac{ \partial \mathbf{X} }{\partial \Theta} \cdot \frac{\partial
  \mathbf{X}}{\partial \Theta} $, etc, and are completely defined by the parametrization given in Equation \ref{eq:X}.  Equation \ref{eq:jacobian_eqn} then provides and equation for the Jacobian on the flux surface.  The remaining components of $ \partial \mathbf{X} / \partial \psi $ are determined by the conditions consistent with the MHD equililbrium conditions \cite{local3D}.
  
\section{Ballooning stability}

The ideal ballooning stability boundary in $s - \alpha$ space for 3-D
local equilibria can then be evaluated.  These marginal stability
boundaries can be constructed for 3-D configurations in a manner
analogous to the traditional axisymmetric procedure \cite{3dballoon}.  The angular
variables are transformed to another set which follows magnetic field lines, $ \eta = \zeta $ (which marks the position along
a magnetic field line) and $ \alpha_{0} = \Theta - \zeta / q $ (the
field line label).  The lowest order incompressible ballooning eigenvalue equation at
is written using $ R_{0} $ and $ \hat{V}^{'} = R_{0} /
B_{0} $ to normalize all physical quantities,
\begin{eqnarray}\label{eq:balloon}
& \frac{\partial}{\partial \eta} \left [ \frac{\hat{V}^{'}}{\sqrt{g}}
  \frac{R_{0}^{4}}{|\nabla \psi|^{2} \hat{V}^{' 2}} ( 1 + \Lambda^{2}
  ) \frac{\partial \xi}{\partial \eta} \right ]  \nonumber\\
& - \frac{\sqrt{g}}{\hat{V}^{'}}\frac{R_{0}^{2}}{\hat{V}^{'} |\nabla
    \psi|} \frac{ \alpha c_{0}^{1/2}}{q^{2}} \left [ R_{0} \kappa_{n} +
    \Lambda R_{0} \kappa_{g} \right ] \xi \nonumber\\
& = - \left [
    \frac{\sqrt{g}}{\hat{V}^{'}} \frac{R_{0}^{4}}{|\nabla \psi|^{2}
      \hat{V}^{' 2} } ( 1 + \Lambda^{2} ) \right ] \omega^{2} \xi .
\end{eqnarray}
Here, $\omega^{2}$ is a normalized eigenvalue, $\kappa_{g}$ and $\kappa_{n}$ are
the geodesic and normal curvatures respectively ($ \mathbf{\kappa} =
\kappa_{n} \hat{n} + \kappa_{g} \hat{b} \times \hat{n} $), and $ \Lambda $ is
the integrated local shear given by
\begin{equation}\label{eq:integrated_shear}
\Lambda = \frac{ |\nabla \psi|^{2} \hat{V}^{\prime 2}}{R_{0}^{4}}
\frac{R_{0}}{\hat{V}^{'} B} \int_{\eta_{0}}^{\eta} d \eta \left [ -s \frac{c_{0}}{q} + \frac{\partial}{\partial \eta}
    D \right ] .
\end{equation}
Here, $\eta_{0}$ plays the dual role of both the radial wave number of the
mode as well as the starting point for field line integration.  The
quantity D is related to the local variation in the magnetic shear and
will be treated in more detail subsequently.  One important feature of ballooning stability
analysis in 3-D (relative to 2-D) is that the ballooning eigenvalue now depends on the field
line label $ \alpha_{0} $.  In this work, separate marginal stability
curves are calculated for different values of $\alpha_{0}$.  At each value of
s and $\alpha$, the ballooning eigenvalue equation is solved for
values of $ \eta_{k} $ spanning one poloidal rotation of the magnetic
surface to find the most unstable (or least stable) eigenvalue.  In
practice, for the 3-D equilibria considered here, the field line labeled by $\alpha_{0} = 0 $ has been found
to be the most unstable field line on any given surface.  

\begin{figure}
\centering
\includegraphics[width=8cm]{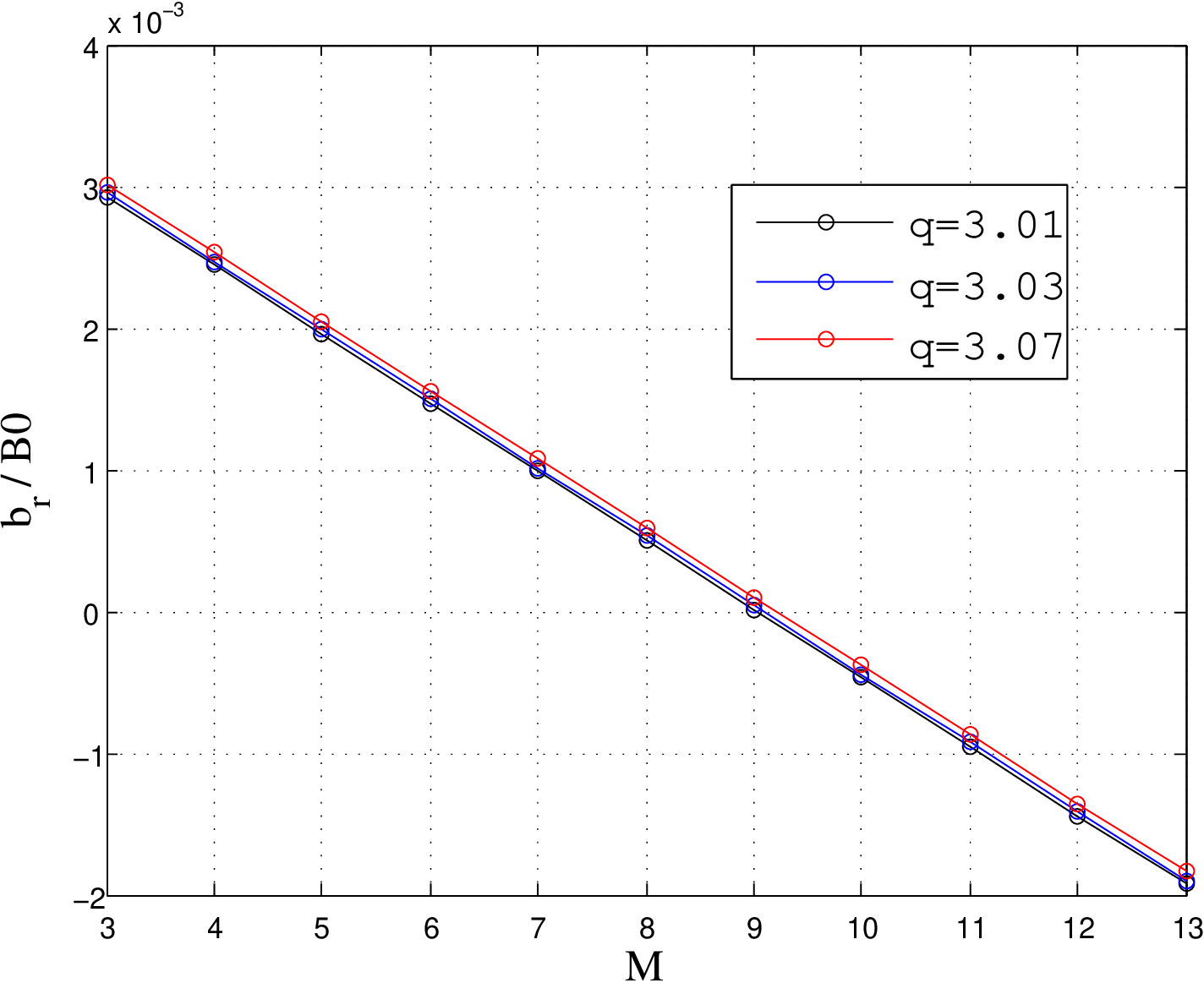}
\caption{Spectrum of the radial magnetic perturbation for 3-D equilibria with several q values where $N=3$ and $\gamma_{i} = 10^{-3}$.}
\label{br}
\end{figure}

The primary focus of this work is on the ideal ballooning stability
properties of a family of equilibria calculated using this method.
An axisymmetric equilibrium with $ A = 3.17 $, $q = 3.03$,
$\kappa = 1.66 $, $\delta = 0.416$, $s_{\kappa} = 0.7 $, $s_{\delta} =
1.37 $, and
$d_{r} R_{0} = -0.354 $ is used as a base case. A set of equilibria are constructed with the same axisymmetric shaping, but with 3-D perturbations with $
\gamma_{i} = 0.001 $ , $ N_{i} = 3 $, and $M_{i} = 4 .. 14 $ added.  The different 3-D equilibria employ a series of q values
approaching 3.  Evaluating Eq. \ref{eq:Br} for these parameters
gives values of $B_{\rho} / B_{0} \leq 3 \times 10^{-3} $ for all harmonics
of the equilibria examined here, with the $ M = 9, N =3 $ component
being of order $10^{-5}$ (or even smaller as q approaches 3).  The spectrum of $ B_{\rho} / B_{0} $ is shown for several of the 3-D 
equilibria in Figure \ref{br}.

\begin{figure}
\centering
\includegraphics[width=9cm]{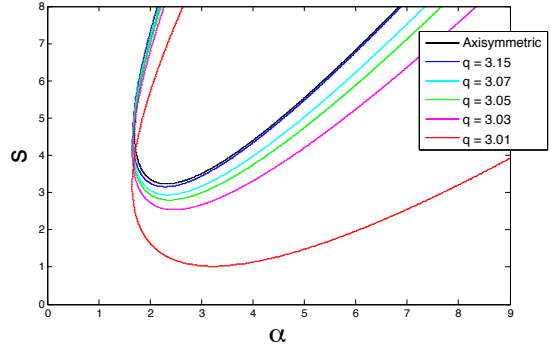}
\caption{Marginal ballooning stability curves for the axisymmetric
  equilibrium with $q=3.03$ and equilibria with 3-D fields added for q
  values 3.15, 3.07, 3.05, 3.03, and 3.01.  All stability curves shown
here are for the field line labeled by $\alpha = 0$, which is
generally the most unstable field line on each surface.}
\label{RMP_boundary}
\end{figure}

The marginal ballooning stability curves for an axisymmetric
equilibrium as well as a family of 3-D equilibria with q ranging from
3.15 to 3.01 are shown in Figure \ref{RMP_boundary}.  Surprisingly, even very weak
3-D fields are capable of appreciably modifying the marginal stability
boundary.  The stability boundary is particularly sensitive to the q
value, with the unstable region expanding as it approaches 3.  This
suggests that there is a resonant mechanism which affects the
stability boundary, which is still weakly operative even when $q=3.15$.  It is also worth noting that the stability boundary expands dramatically at $q=3.01$, which we will return to subsequently.  

\begin{figure}
\centering
\includegraphics[width=8cm]{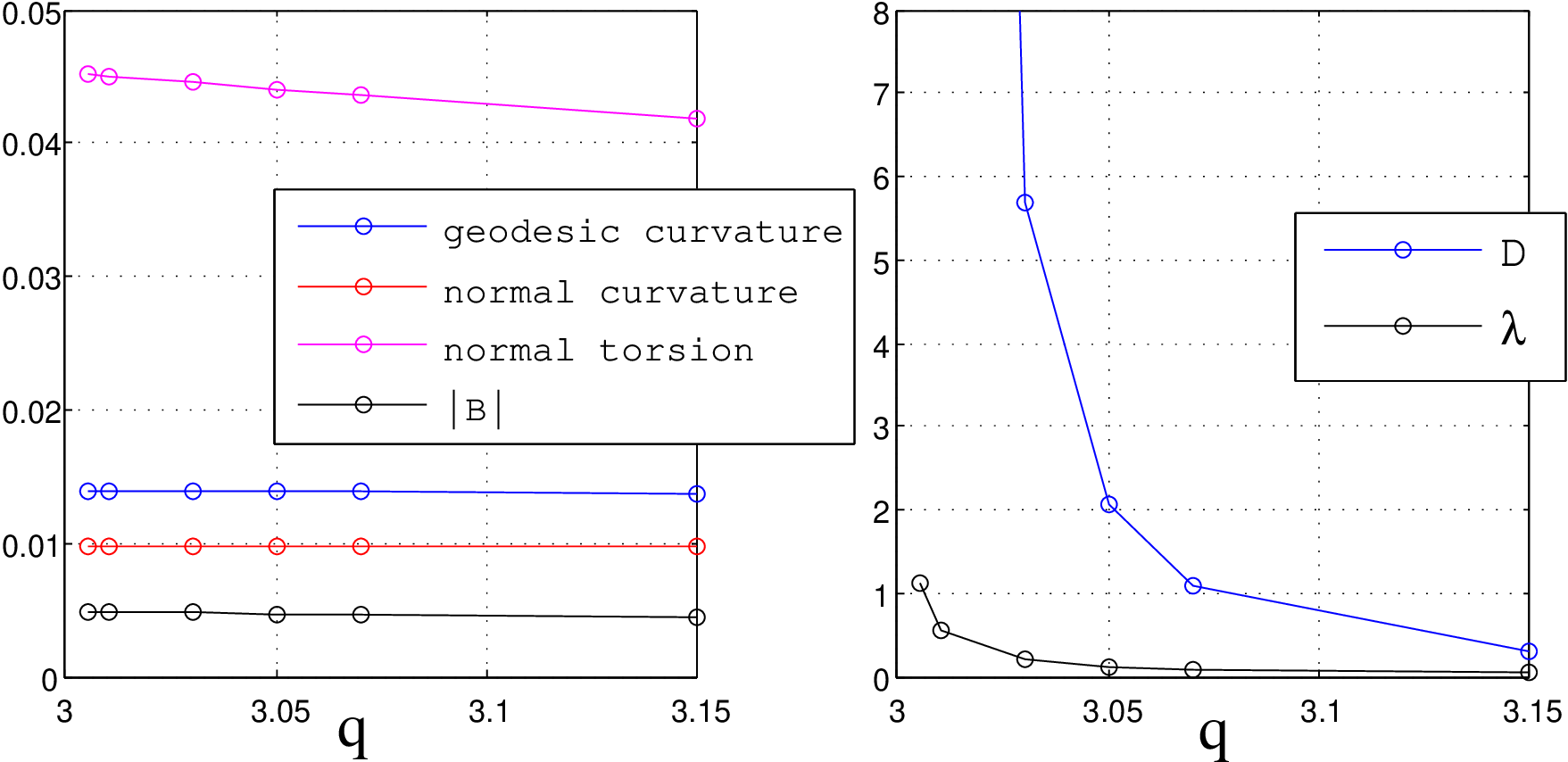}
\caption{A measure of the degree to which the 3-D deformation modifies various quantities.  
For some quantity Q, evaluated on a real space grid, the following is calculated: $ \left [ \sum_{i,j} ( Q_{ij}^{3D} - Q_{ij}^{2D} )^{2} \right ]^{1/2} / \left [ \sum_{i,j} ( Q_{ij}^{2D} )^{2} \right ]^{1/2} $.  A value of $0.01$ corresponds to an average perturbation on the order of $1 \%$.  }
\label{norms}
\end{figure}

The important question is then, how can a set of magnetic perturbations with a
magnitude on the order of
$ 10^{-4} $ of the background field have such a dramatic impact on the
ballooning stability boundary?  We begin by inspecting all the terms appearing in Eq.
(\ref{eq:balloon}).  A measure of how strongly various quantities are perturbed by the 3-D
deformation is shown in Figure \ref{norms}. The 3-D perturbation has no appreciable affect on $
B^{2} , | \nabla \psi |^{2}, \kappa_{g}, \kappa_{n} $ or $ \sqrt{g}
$.  However, the integrated local shear, $ \Lambda $, is substantially
modulated.  By examining Eq. (\ref{eq:integrated_shear}) it is
clear that this modulation is due the local variation of the magnetic
shear within the surface, encapsulated in the quantity D.

The local variation of the magnetic shear is determined by a magnetic
differential equation, 
\begin{eqnarray}\label{eq:D}
\left ( \frac{\hat{V}'}{\sqrt{g}} \right ) \frac{\partial } {\partial
  \eta } D = - s \frac{c_{0}}{q} \left [ \frac{B^{2} R_{0}^{2} }{ | \nabla \psi
  |^{2}} \frac{1}{ c_{0} } - \frac{\hat{V}^{\prime}}{\sqrt{g}} \right ] \nonumber\\
- \alpha \frac{c_{0}^{1/2}}{2 q^{2} } \left [ \frac{B^{2} R_{0}^{2}}{ | \nabla \psi |^{2}}
\frac{\lambda}{\hat{V}'} - \frac{B^{2} R_{0}^{2}}{ | \nabla \psi
  |^{2}} \frac{ \left < \frac{B^{2} R_{0}^{2}}{| \nabla \psi |^{2}} \frac{
    \lambda } {\hat{V}'} \right > }{c_{0}} \right ] \nonumber\\
+ 2 \left [ \frac{B^{2} R_{0}^{2}}{ | \nabla \psi |^{2}} \frac{ \left < \frac{
    B^{2} R_{0}^{2}}{| \nabla \psi |^{2}} \tau_{n} R_{0} \right >}{c_{0}} -
\frac{B^{2} R_{0}^{2}}{| \nabla \psi |^{2}} \tau_{n} R_{0} \right ] .
\end{eqnarray}
The right hand side of this expression can be split into three component
parts relating to different physical effects.  The first term on the
right hand side relates to local shearing of the magnetic field
due to changes in the average magnetic shear.  The second term relates to shearing of the magnetic field
due to Pfirsch-Schlüter currents.  The parallel current is written
$J_{\parallel} / B = < \mathbf{J} \cdot \mathbf{B} > / < B^{2} >
+ p ' \lambda $ where the quantity $\lambda$ is calculated consistent
with the quasi-neutrality condition $ \nabla \cdot \mathbf{J} = 0 $.  The third term
relates to shearing of the magnetic field due to the geometric
properties of the surface (through the normal torsion $\tau_{n} = -
\hat{n} \cdot ( \hat{b} \cdot \nabla ) ( \hat{b} \times {\hat{n}} ) $).
The normal torsion is completely described by $\mathbf{X} ( \Theta,
\zeta ) $ \cite{local3D}. Examining these three component parts shows that the 3-D perturbation
does not appreciably affect the first term.  There is a modest 
effect on the normal torsion but the dominant modulation of the local
shear comes from the Pfirsch-Schlüter currents.  

As a point of clarification, we note 
that the quantity D appears in the integrated local magnetic shear (Equation \ref{eq:integrated_shear}),
however it is $\partial D / \partial \eta$ (the left hand side of Equation \ref{eq:D}) which appears in the
local magnetic shear.  The resonant amplification of the right hand side of the Pfirsch-Schlüter equation
occurs once more when calculating D - this is due to the fact that the helical modulation of the local magnetic 
shear is nearly pitch resonant with magnetic field lines.  Therefore, the integrated local magnetic shear (which
explicitly appears in Equation \ref{eq:balloon}) is modulated much more than the local magnetic shear.  However, 
for the remainder of this work we will focus on the local magnetic shear, $\partial D / \partial \eta $, as it plays a more intuitive role in 
ballooning stability.  

\section{Local shear modulation}

\begin{figure}
\centering
\includegraphics[width=6.5cm]{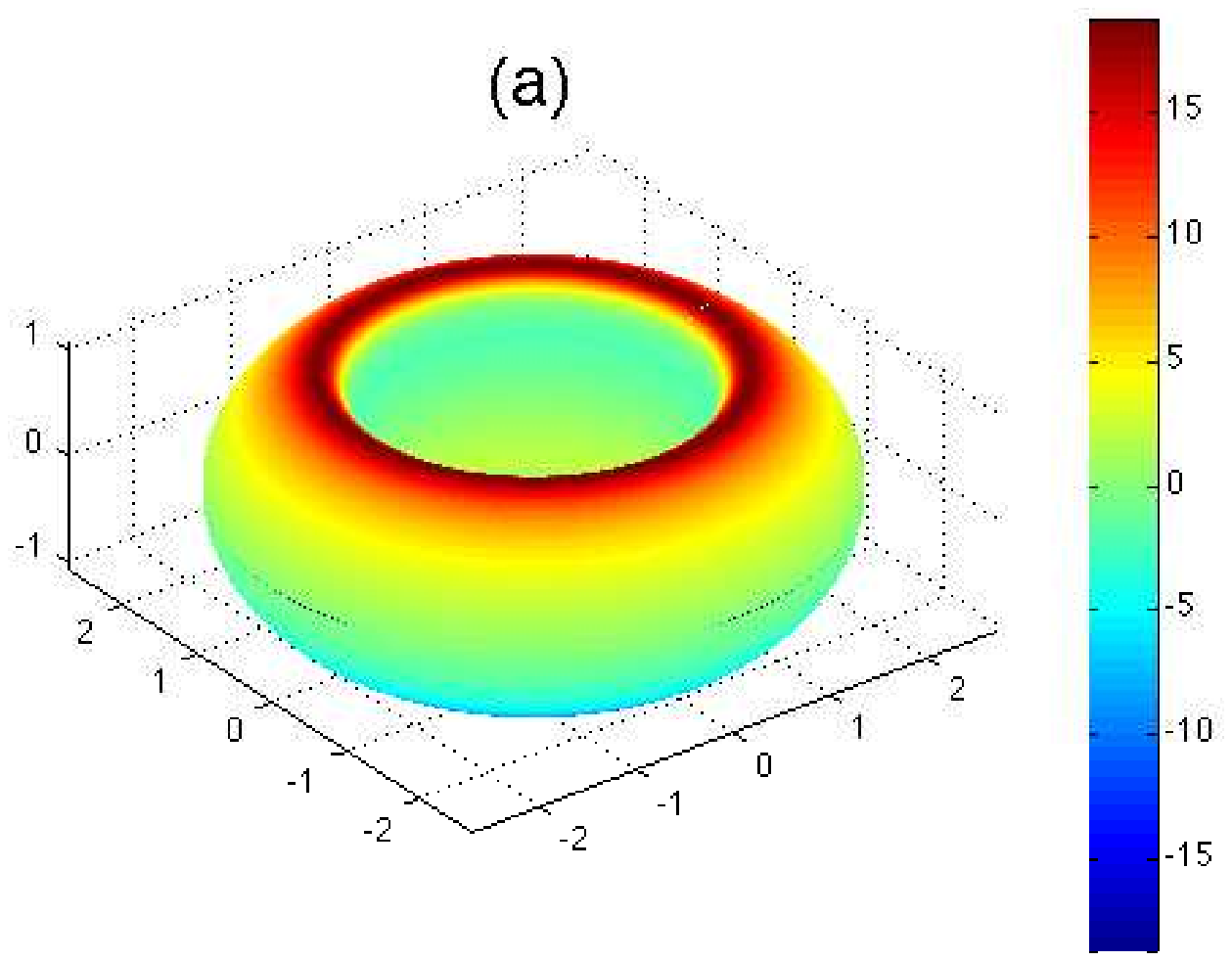}
\includegraphics[width=6.5cm]{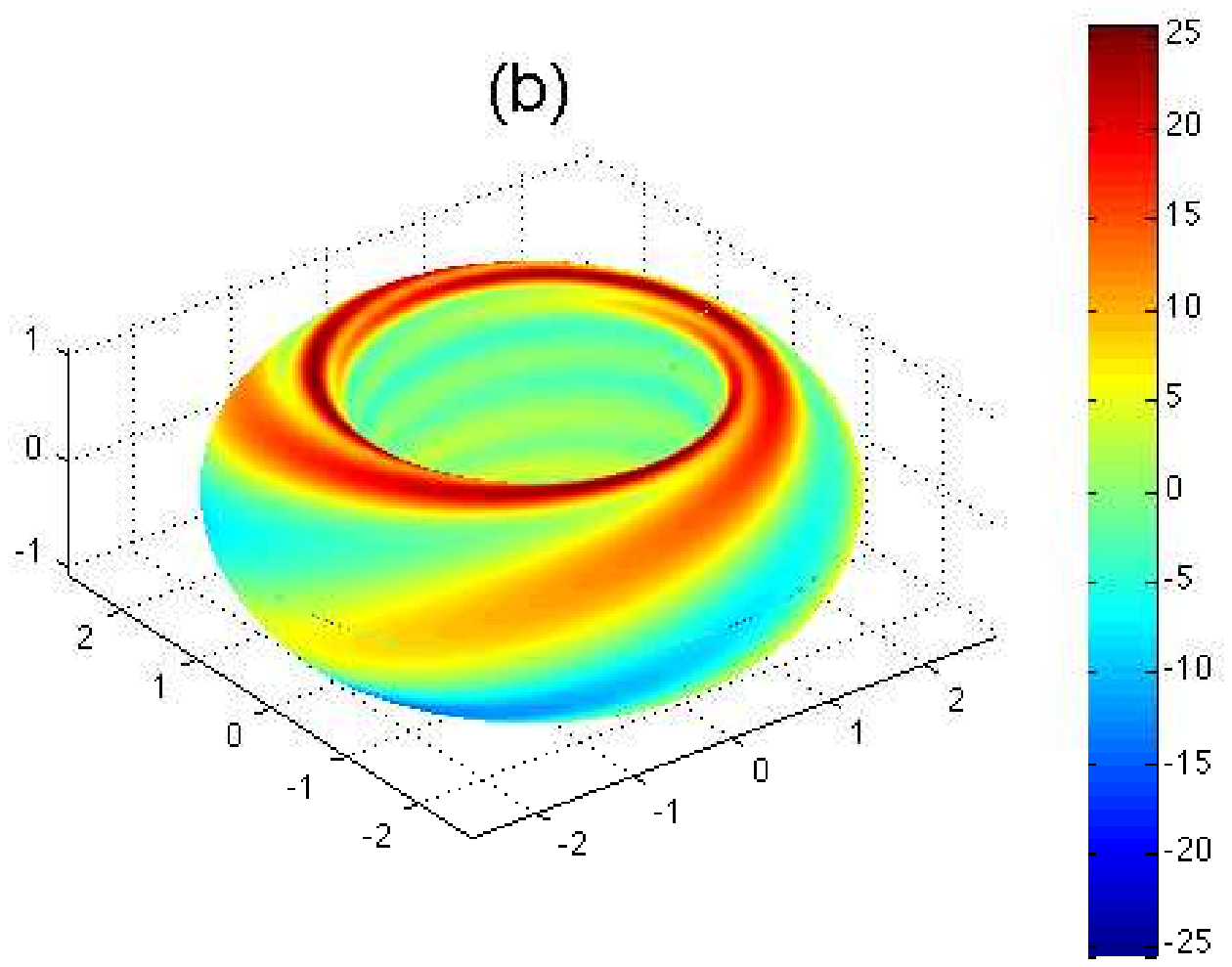}
\includegraphics[width=6.5cm]{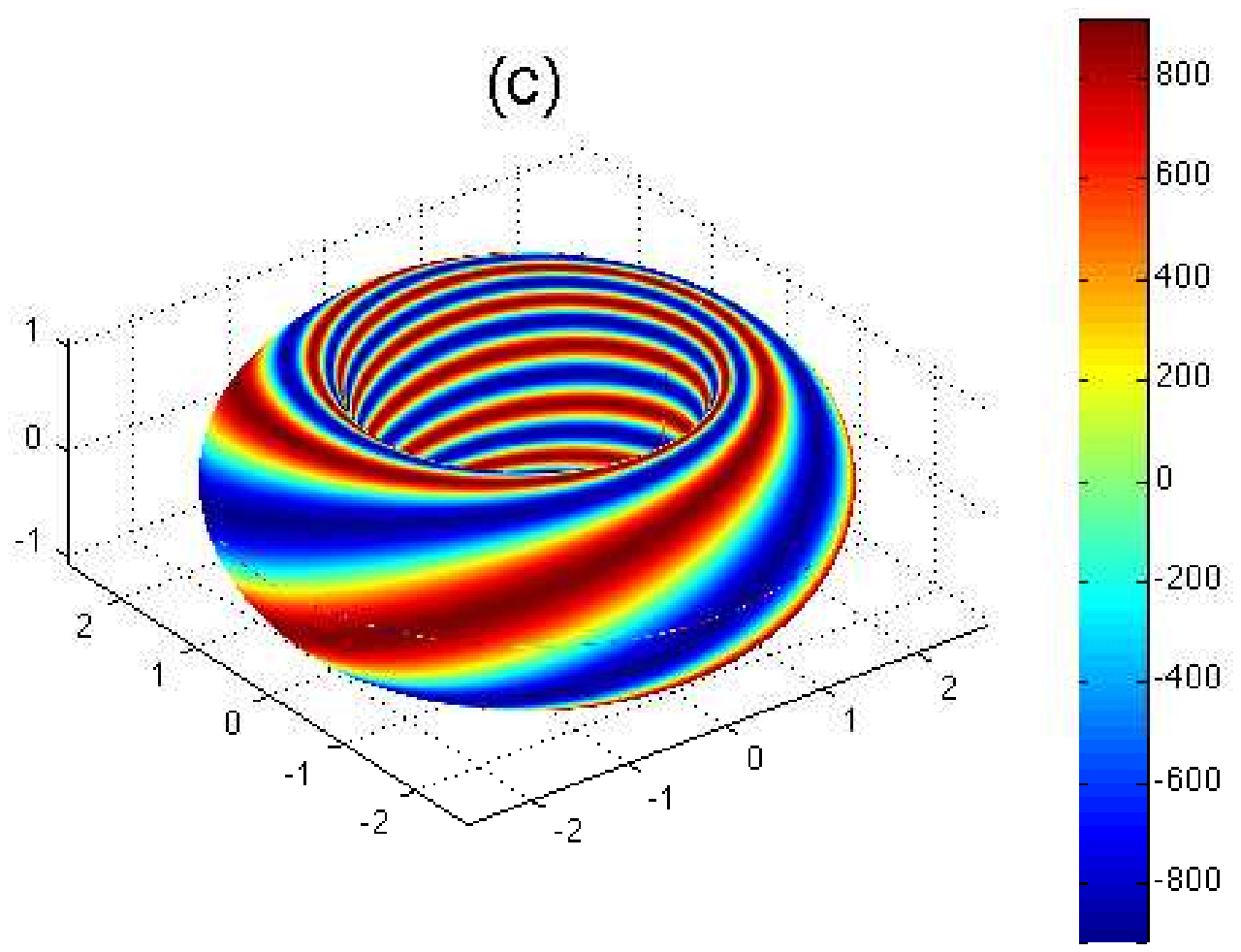}
\caption{Contours of the quantity D on the magnetic surface for (a) the axisymmetric
  equilibrium with $q=3.03$, and equilibria with 3-D fields added and q
  values of (b) 3.15 and (c) 3.01.}
\label{3d_localshear}
\end{figure}

The Pfirsch-Schlüter current spectrum is determined by a magnetic
differential equation,
\begin{equation}
\left ( \frac{\hat{V}'}{\sqrt{g}} \right ) \frac{\partial}{\partial \eta} \frac{
  \lambda } { \hat{V}' } = 2 \mu_{0} \kappa_{g} \frac{ | \nabla \psi |
}{B} .
\end{equation}
Near rational magnetic surfaces, very small perturbations to the
quantities on the right
hand side of this equation can dominate the local magnetic shear.  Using a
Fourier decomposition, the Pfirsch-Schlüter harmonics are now given by
$\lambda = \sum_{M N} \lambda_{M N} cos ( M \Theta - N \zeta )$ with
\begin{equation}\label{eq:lambda_fourier}
\lambda_{M N} =  \frac{1}{-N + M / q} \left ( 2 \mu_{0} \kappa_{g} \frac{ |
  \nabla \psi |}{B} \sqrt{g} \right )_{M N} .
\end{equation}
The differential operator in these magnetic differential equations
becomes singular as q approaches a rational value, which can be seen
as the denominator on the right hand side approaching zero.  Even if
the resonant component of the radial magnetic field is perfectly
shielded, non-resonant components of the radial magnetic field can easily
result in large responses to the right hand side of Eq.
(\ref{eq:lambda_fourier}).  Figure \ref{3d_localshear} shows the
quantity D for several of the equilibria analyzed here.  As q
approaches 3, the peak amplitude of the variation of the local shear
is greatly enhanced and the structure becomes dominantly helical. 

\begin{figure}
\centering
\includegraphics[width=6.5cm]{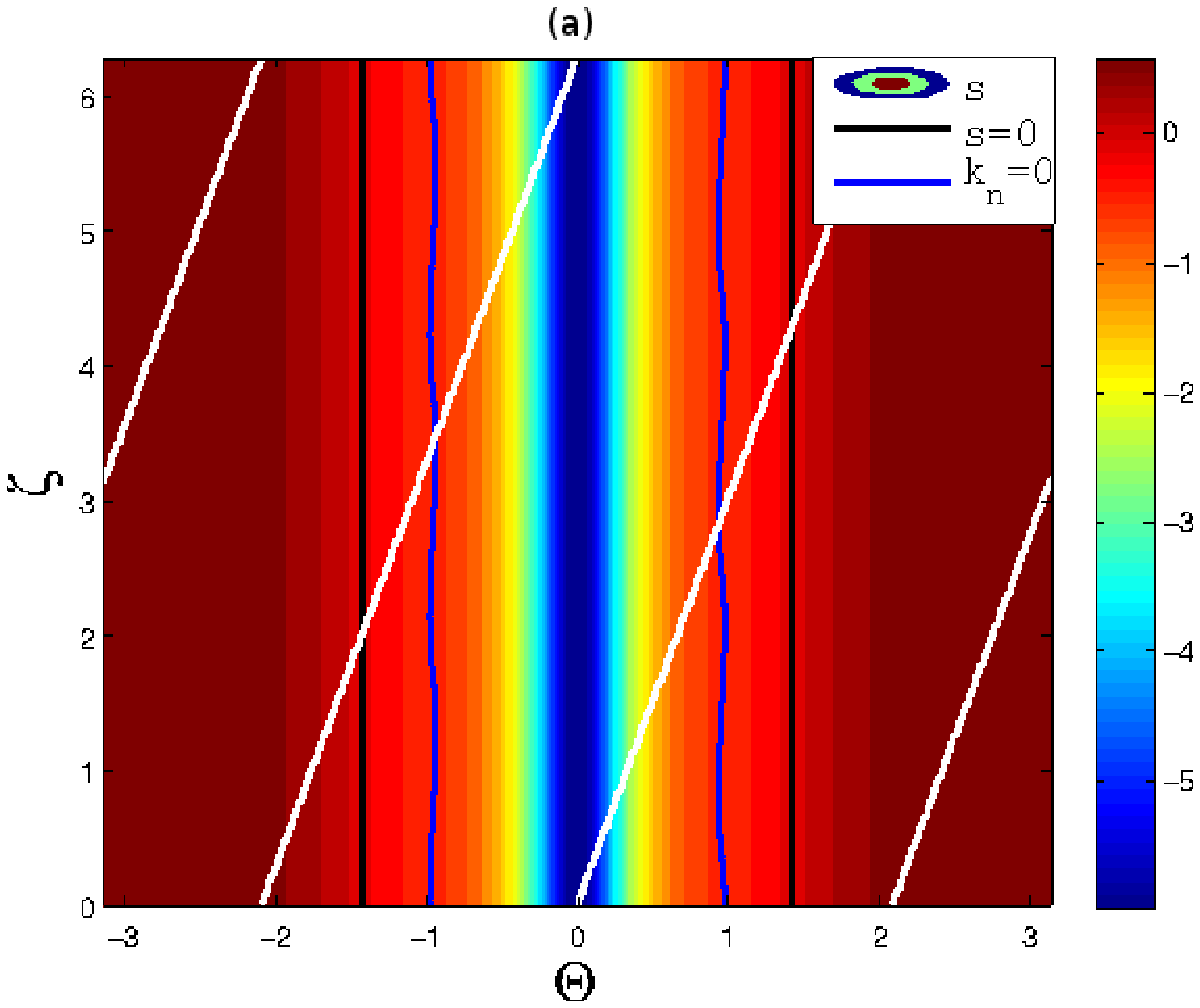}
\includegraphics[width=6.5cm]{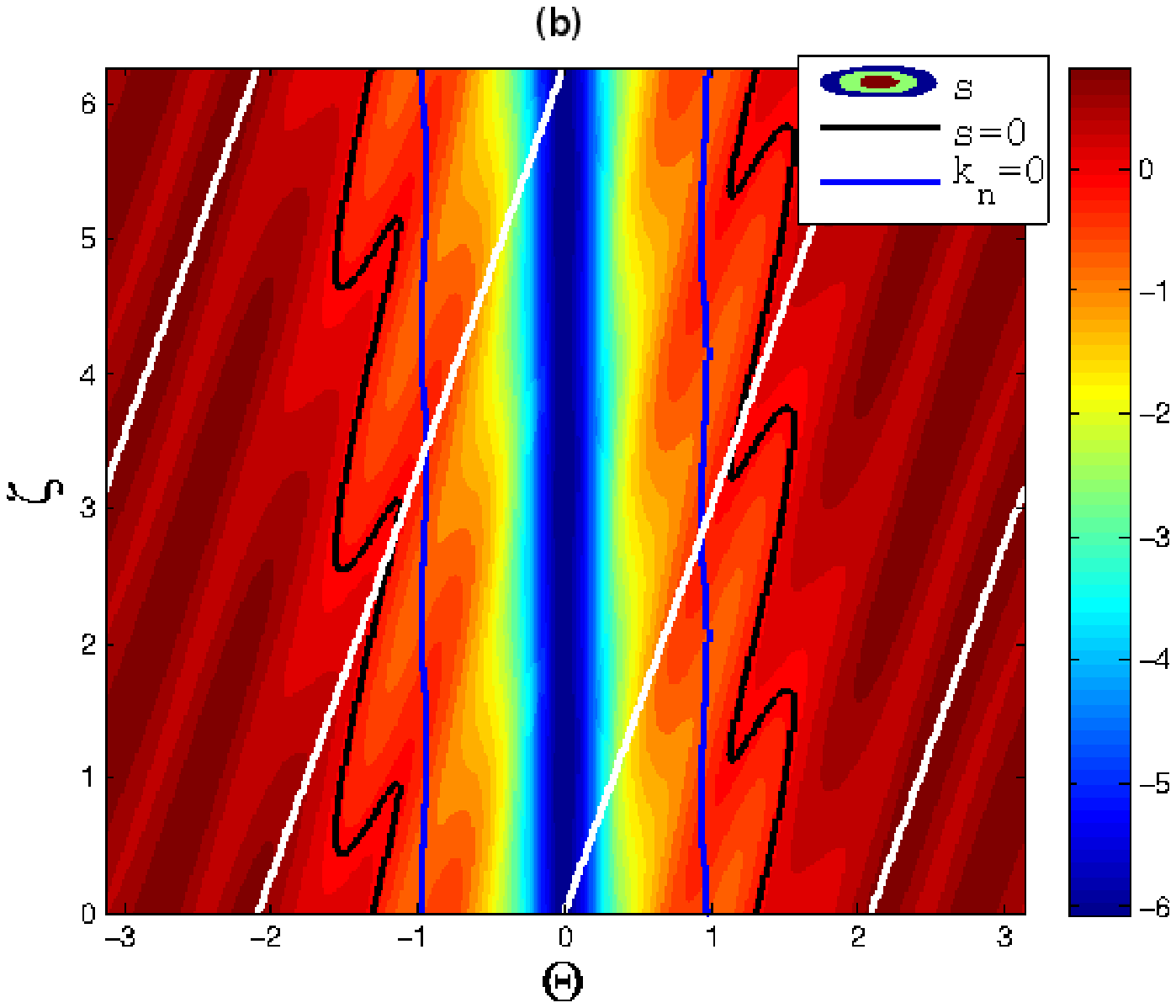}
\includegraphics[width=6.5cm]{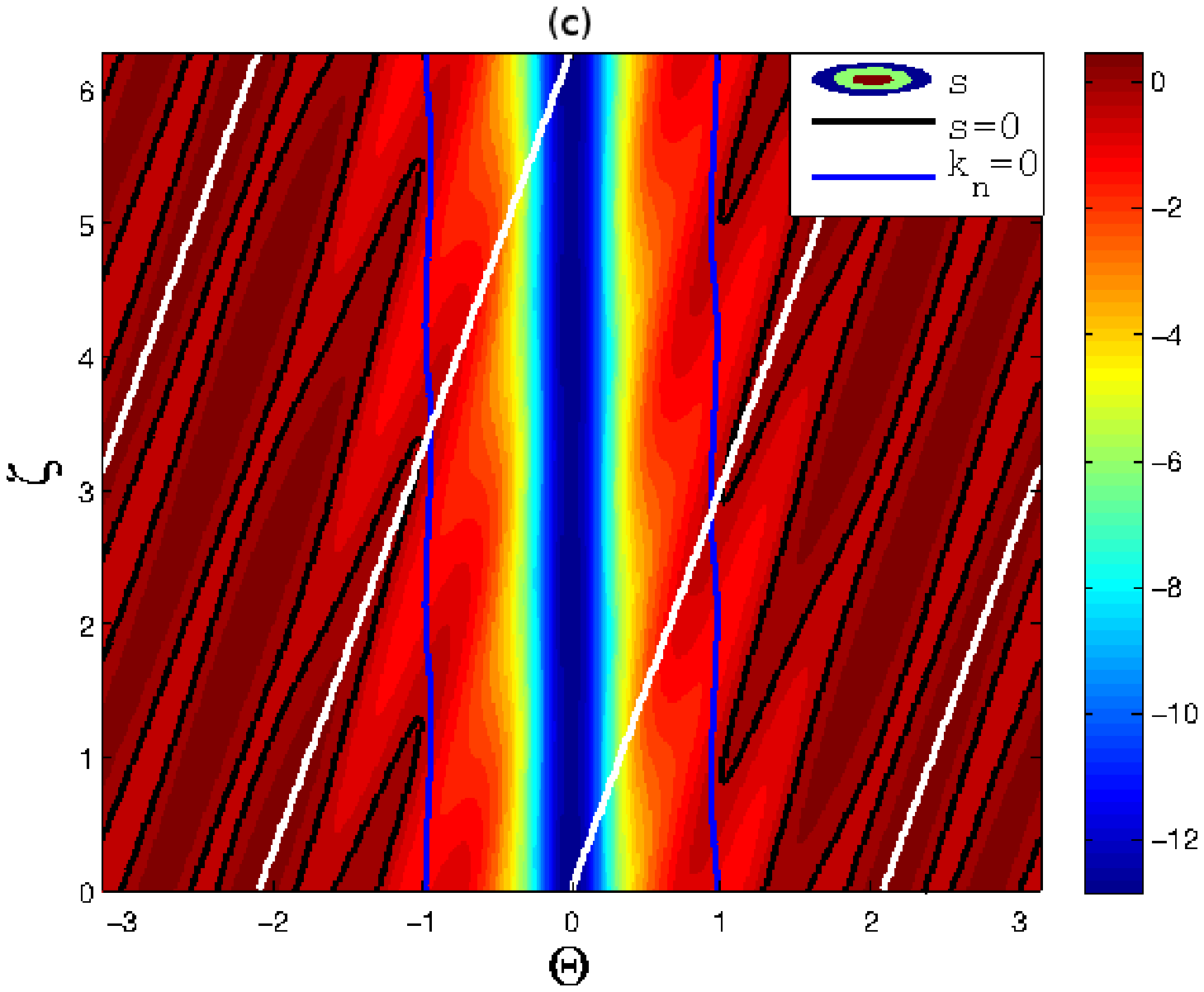}
\includegraphics[width=6.5cm]{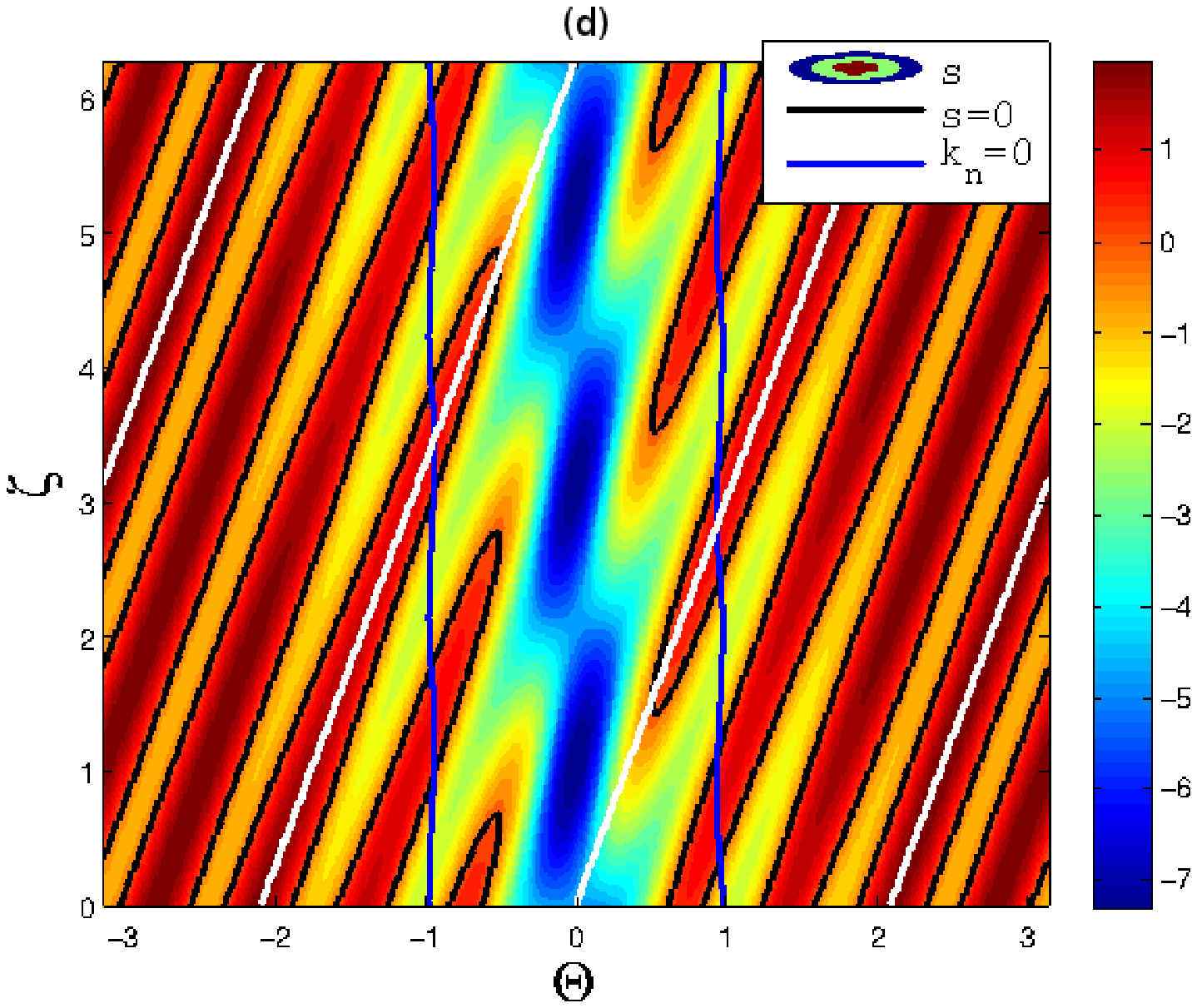}
\caption{Contours of the local magnetic shear for (a) the axisymmetric base case,
and 3-D equilibrium with (b) $q=3.07$, (c) $q=3.03$, and (d) $q=3.01$.  The thick black line marks
the zero contour of the local magnetic shear, the thick blue line marks the zero contour of the 
normal curvature (which is negative near $\Theta=0$), and the thick white line shows the path of a magnetic field line
which passes through ($\Theta=0 , \zeta=0$).}
\label{local_shear_contours}
\end{figure}

In order for ballooning modes to be driven unstable, the local magnetic 
shear must have a sufficiently low magnitude in regions of negative normal curvature.  
Figure \ref{local_shear_contours} shows the distribution of the local magnetic shear for several 3-D equilibria with $s=1$ and $\alpha=3$, where the $q=3.01$ equilibrium is near marginality but all other equilibria are well within the stable region.  In the axisymmetric equilibrium, the local magnetic shear reaches relatively large
negative values at the outboard midplane where the normal curvature is negative.  The 
local magnetic shear passes through zero in a region of positive normal curvature.  
As q approaches 3, the helical modulation shifts the null in the local magnetic shear 
towards the outboard midplane.  At $q=3.01$, where the ballooning stability boundary is
quite dramatically expanded, the null in the local shear overlaps with the region of 
negative normal curvature.  This is strongly conducive to ballooning instability and
explains the expansion of the unstable region in $s-\alpha$ space.

\begin{figure}
\centering
\includegraphics[width=8cm]{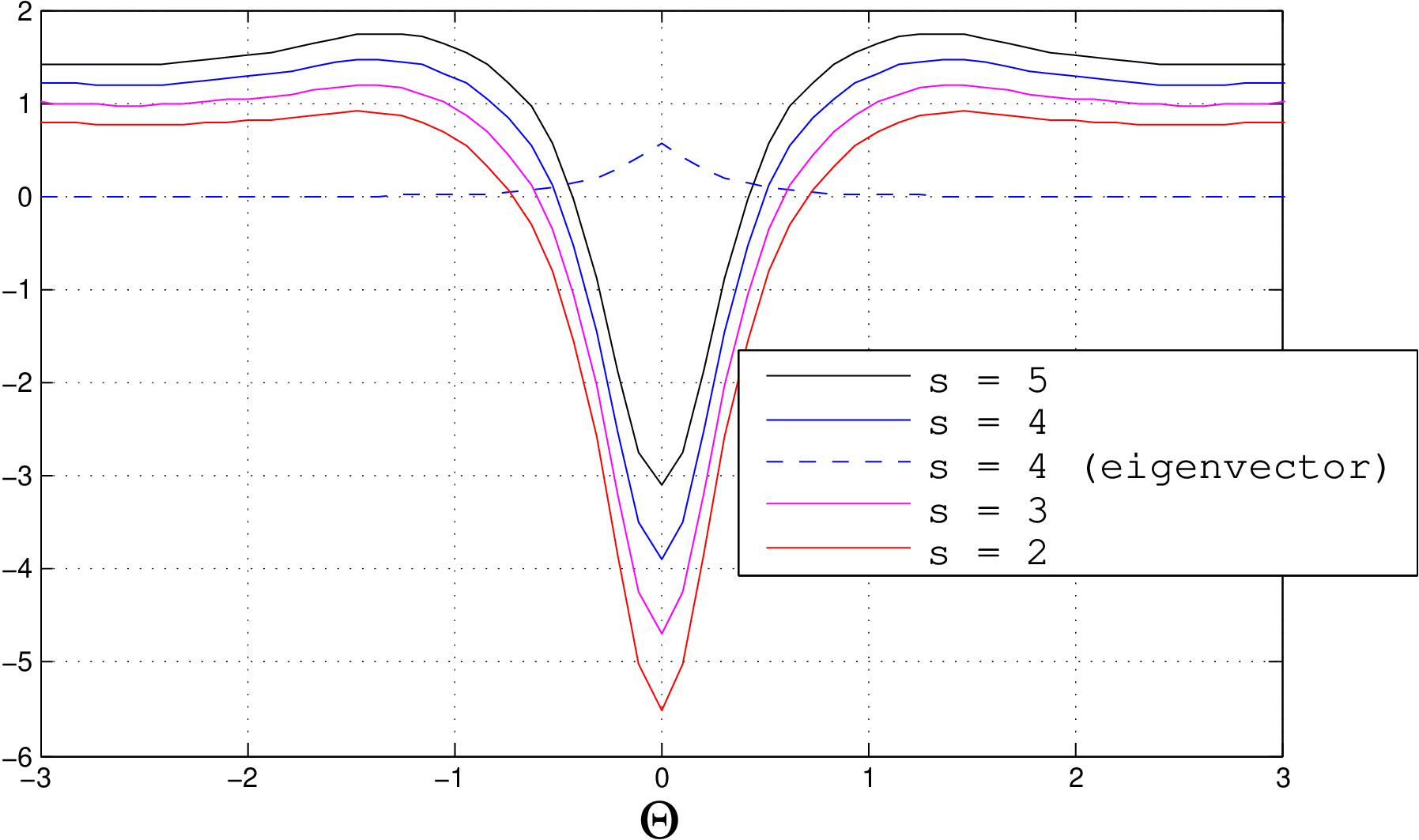}
\caption{3-D equilibrium with $q=3.05$, with $\alpha=4$.  The local magnetic shear is shown
at different values of the surface averaged shear.  The ballooning eigenvector is also shown
for the unstable $s=4$ case.  Here, the point of marginal stability is at a surface averaged shear of $\sim 3.75$.  
In all cases, $\alpha_{0}=0$.}
\label{shear_stabilization}
\end{figure}

If we limit our attention to the field line which passes through $\zeta=0$ and leave $\eta_{0}$ fixed at 0, 
which is generally the most unstable point on a surface, more insight can be gained.  First we
examine the 3-D equilibrium with $q=3.05$ at $\alpha=4$ and look at the effect of varying the
surface average magnetic shear.  Figure \ref{shear_stabilization} shows the structure of the 
local magnetic shear as the flux surface averaged shear is raised from 2 to 5 (the stability boundary is
at roughly 3.75), and the unstable eigenvector at the average shear of 4.  As the surface averaged
shear is raised, the strongly negative magnetic shear at the outboard midplane ($\Theta=0$) is weakened.  
The zeros in the local magnetic shear also move closer to the outboard midplane, where the unstable
eigenvector peaks.  

\begin{figure}
\centering
\includegraphics[width=7.5cm]{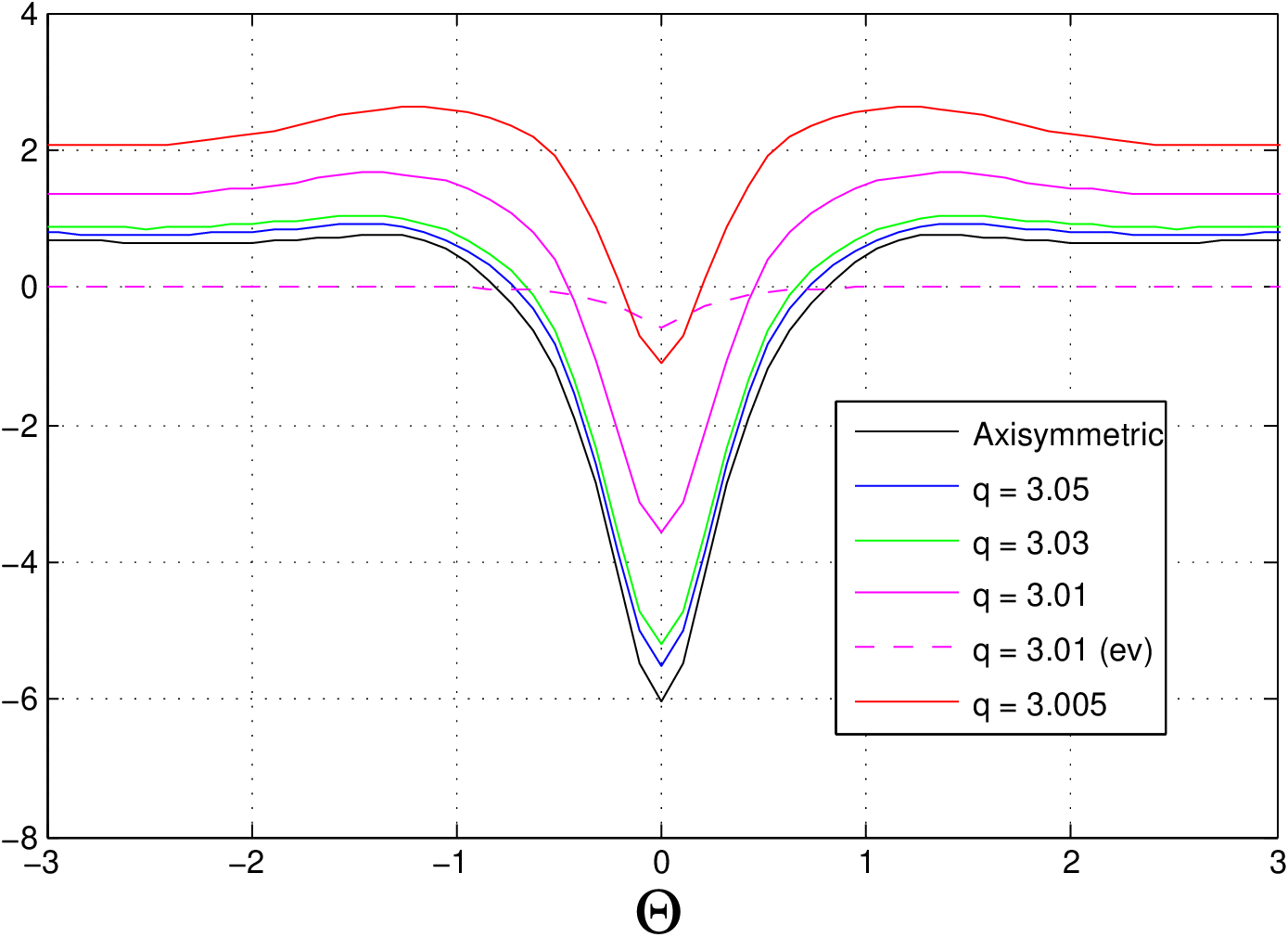}
\caption{The local magnetic shear for various equilibria with $s=2$ , $\alpha=4$ , $\alpha_{0}=0$.  The
unstable eigenvector is also shown for the $q=3.01$ equilibrium.  A 3-D equilibrium with $q=3.005$ is also shown
here to demonstrate how strong the local shear modulation becomes near resonance.}
\label{3d_stabilization}
\end{figure}

This same process can be observed by examining equilibria with different q values at 
($s=2$,$\alpha=4$), where all equilibria except $q=3.01$ are stable.  Figure \ref{3d_stabilization}
shows the structure of the local magnetic shear for several equilibria, and
the unstable eigenvector at $q=3.01$.  The local magnetic shear for an equilibrium with $q=3.005$
is also shown (this equilibrium is so unstable that the stability boundary does not even fit in Figure \ref{RMP_boundary}.)
As the safety factor approaches the resonant value, the local magnetic shear is helically modulated across the surface.
For some magnetic field lines, this is catalytic for ballooning instability due to the lower magnitude of the
local magnetic shear at the outboard midplane, and the shift in the zeros of the local magnetic shear towards the outboard midplane.    

\begin{figure}
\centering
\includegraphics[width=8cm]{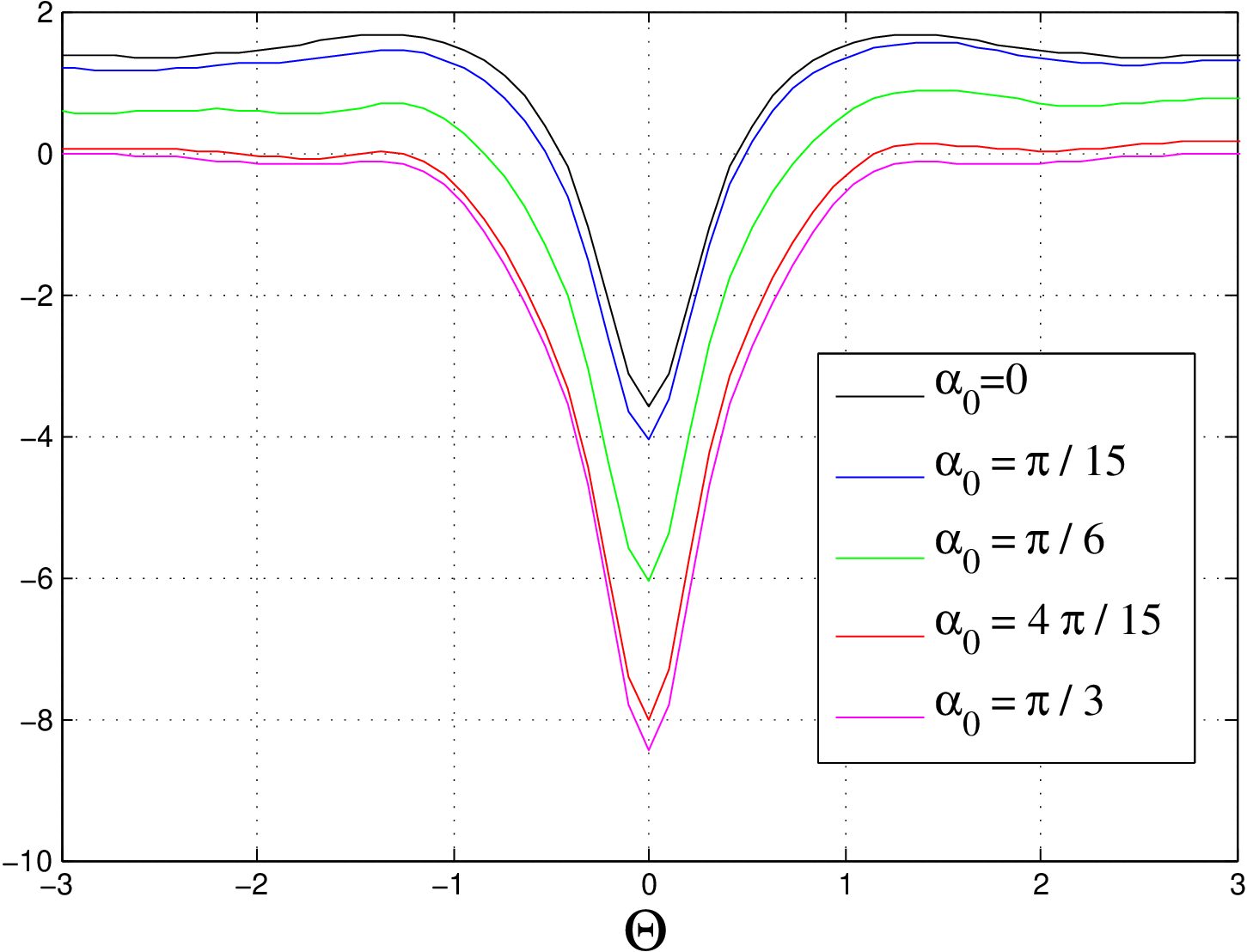}
\caption{Change in local magnetic shear with field line label for the 3-D equilibrium with $q=3.01$, $\alpha=4$, $s=2$.  The field lines
labeled $\alpha_{0}=0$ and $\alpha_{0}=\pi/15$ are unstable while the other 3 are stable.  The 3-D perturbation is destabilizing for some field lines but stabilizing for others.}
\label{s_vs_alpha0}
\end{figure}

\begin{figure}
\centering
\includegraphics[width=8.5cm]{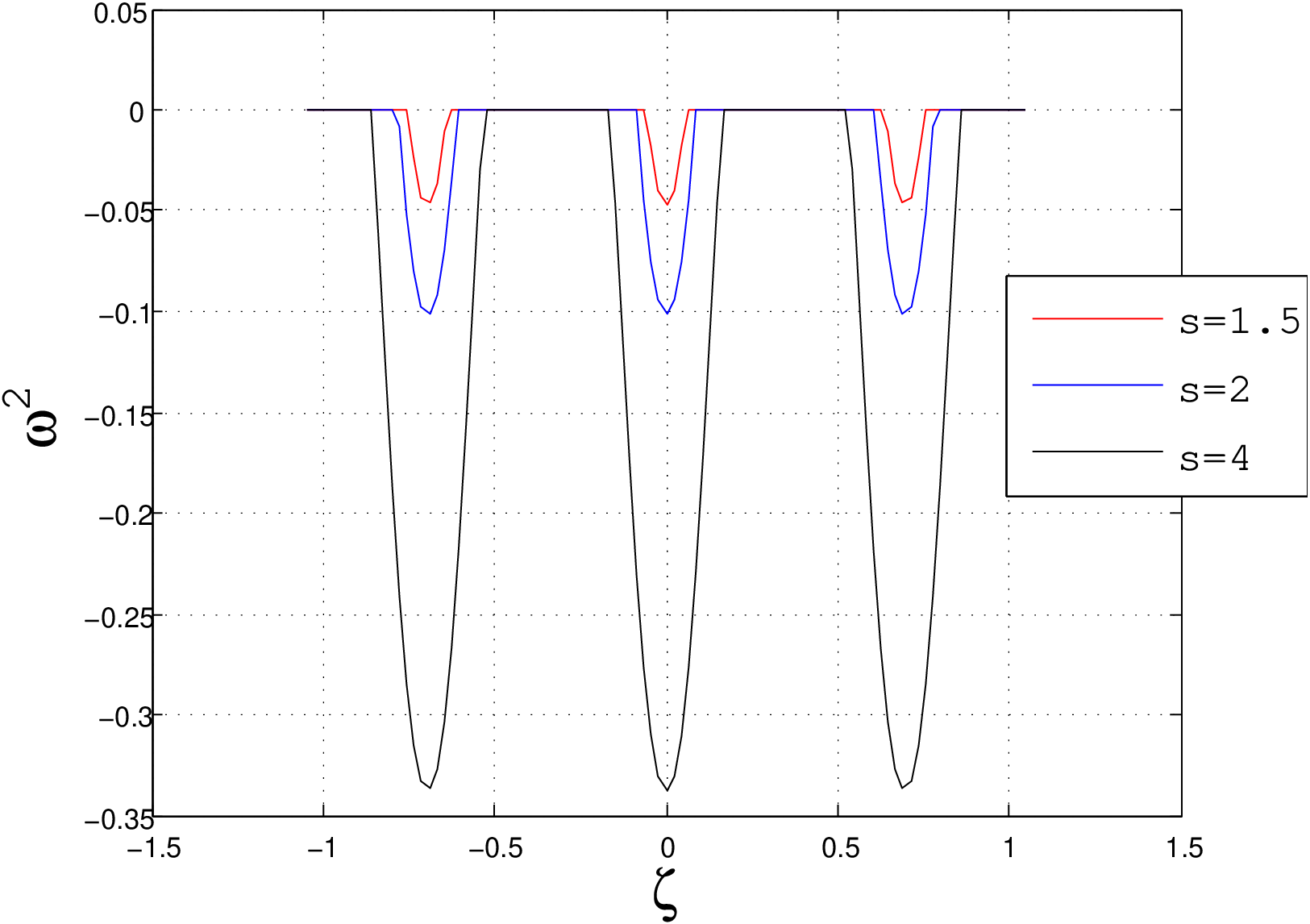}
\caption{The ballooning eigenvalue as a function of the field line label for the 3-D equilibrium with $q=3.01$ at several
values of the surface averaged shear.  Here again, $\eta_{0}$ was varied to keep the computational domain centered at 
$\Theta=0$.  $s=1.5$ is close to marginality and the $s=2,4$ calculations go deeper into the unstable region.  Even at
$s=4$, a portion of the magnetic field lines are still stable.}
\label{alpha0_variation}
\end{figure}

However, for some of the magnetic field lines on the surface the result of the
3-D perturbations is to draw the zeros of the local magnetic shear away from the 
outboard midplane.  Therefore the 3-D perturbations have a destabilizing effect
on some areas of the surface but a stabilizing effect on others.  Figure \ref{s_vs_alpha0}
shows the change in the structure of the local magnetic shear as the field line label is varied
($\eta_{0}$ is also varied such that the computational domain is centered at the outboard midplane).  
As the toroidal angle is varied away from $\zeta = 0$, the magnitude of the local magnetic shear
is strengthened at the outboard midplane and the zeros shift towards the inboard side.  A scan
of the ballooning eigenvalue with respect to the toroidal angle is shown in Figure \ref{alpha0_variation},
showing that even deep within the unstable region, some portion of magnetic field lines are still
ballooning stable.  

In principle, for 3-D configurations the stability of the entire flux surface should be studied 
self consistently.  The infinite-n stability analysis is suitable in a tokamak where every magnetic field
line on a surface has the same properties.  However in 3-D geometry, as we have seen, different
field lines can have radically different stability properties, so the analysis of a single field line
may not be sufficient.  In future work we will test the infinite-n MHD ballooning predictions with
gyrokinetic KBM calculations which consider the full surface self-consistently.  

To conclude the stability analysis, we remark that the mechanism by which the local shear modulation
affects ballooning stability is surprisingly simple.  The helical perturbation to the local magnetic shear is nearly 
aligned with magnetic field lines, so an infinite-n ballooning mode located on a particular field line will
mostly experience a uniform increase or decrease in the local magnetic shear (relative to 2-D) when the 3-D perturbation is added.
This can have a stabilizing or destabilizing effect, for the exact same reasons that changing the surface averaged magnetic shear
can be stabilizing or destabilizing depending on the location in $s-\alpha$ space.  We have identified and elucidated this mechanism,
however to perform quantitative predictions will require actual Kinetic Ballooning Mode calculations which consider an entire
flux surface self-consistently, which is planned as future work.

\section{Summary and Discussion}

These calculations demonstrate two important points concerning the use
of externally applied, 3-D RMP fields.  The first is that very small
3-D magnetic perturbations, of experimentally relevant magnitudes, can
drive substantial non-axisymmetric modulation of the Pfirsch-Schlüter
current spectrum and local magnetic shear even when shielding physics 
prevents the presence of magnetic island formation.  This effect is caused by a
pitch resonance between magnetic field lines and the 3-D
geometric deformation of the surface, in particular through a product
of the geodesic curvature and magnetic field strength.  This is
related to the same physical effect which leads to the singular
current problem of 3-D global MHD equilibrium calculations.  

A second point is that this 3-D modulation of the local shear has a significant effect on the ideal ballooning
stability boundaries.  These calculations suggest that the onset conditions for KBM instability can be significantly
lowered near rational surfaces in the presence of 3-D perturbations.  The initial variant of the predictive pedestal model EPED (EPED1) described the onset of KBM turbulence using the approxmation that the marginal stability boundary was of the shape $\alpha_{crit} \sim 1 / s^{1/2}$ \cite{snyder3}, suggesting that pedestal evolution is particularly sensitive to the lower left portion of the stability boundary, which is where it is most strongly perturbed by the 3-D local shear modulation.  
Gyrokinetic modeling of pedestal evolution has found that the inward advance of the pedestal occurs when unstable KBMs at the pedestal top are pushed toward marginal stability by the lowering of the surface averaged magnetic shear associated with growing boostrap currents as the pressure gradient rises \cite{dickinson2}.  Given that the dominant effect of the 3-D local shear modulation is to allow instability at much lower surface averaged shear, as seen in Figure \ref{RMP_boundary},  the 3-D pertubations could halt the inward advance of the pedestal by disrupting this stabilization process at the pedestal top once it reaches a low order rational surface.  In this event, type-I ELMs can be avoided.

There are several important consequences for future modeling work.  Studies of plasma microinstabilities in the presence of 3-D RMP fields which utilize 2-D MHD equilibria and include the RMP field separately will not capture the Pfirsch-Schlüter physics.  Proper analysis of this effect requires 3-D MHD equilibrium calculations with a finite pressure gradient at the surface(s) of interest.  The ideal MHD model exhibits a singular parallel current at rational surfaces which in principle must be resolved using a more detailed physics model.  We have sidestepped this by using radially local equilibrium calculations and merely examining 'near-resonant' effects.  However, a better understanding of the nature of the plasma response to 3-D deformation at rational surfaces is desirable.  

In summary, we have presented a model for enhanced transport in the edge region of H-mode tokamaks in the presence
of RMP fields.  Small 3-D distortions of the flux surface shape can lead to helical Pfirsch-Schlüter currents which substantially modulate the local magnetic shear, a quantity which plays an important role in stability of localized modes and their associated anomalous transport.  Even when the resonant components of the radial magnetic perturbation is shielded,
this effect is still operative.   The Pfirsch-Schlüter current and local shear are determined by magnetic differential equations which becomes singular near rational values of $q$.  At a rational surface this singular response will be resolved by physical mechanisms beyond the scope of ideal MHD.  However the near-resonant response is quite significant, and therefore 3-D magnetic perturbations with a broad poloidal mode spectrum may be able to halt the inward evolution of the pedestal as it approaches a low order rational surface by adversely affecting KBM turbulence.

Future work will use linear and nonlinear gyrokinetic calculation of KBMs to test the ideal MHD ballooning prediction.  A better understanding of what physical processes resolve the singular Pfirsch-Schlüter currents present in ideal MHD is desirable to assess the validity of the local 3-D MHD equilibrium calculations as a rational surface is approached.  Sensitivity studies of these results with respect to external magnetic perturbation strength
and spectrum content also need to be pursued.  However, we have demonstrated that local magnetic shear modulation driven by helical Pfirsch-Schlüter currents provides a possible explanation, which does not rely on stochasticity, for the observed enhanced plasma transport during some ELM suppression experiments.

\appendix

\section{Analysis in the shifted circle limit}
\setcounter{equation}{0}
\renewcommand{\theequation}{A\arabic{equation}}

In order to obtain gain more analytic insight into how the 3-D fields
alter the ballooning stability properties, analysis of the ballooning equation
for a particular 3-D equilibrium  can be performed.    
We start from an axisymmetric equilibrium with large aspect ratio, low $\beta$, circular concentric flux surfaces and locally steepened pressure gradient  \cite{cht}.  To this equilibrium, 3-D
fields are added as described in Section II.   
Using Eqs. (1), (2) and (8) to define the equilibrium in the $A \gg 1, \delta = \kappa
= s_{\kappa} = s_{\delta} = d_rR_0 = 0, \gamma_i/r \ll 1$ limit yields asymptotic
estimates for various geometric quantities.  In particular, the normal and 
geodesic curvatures are given by
\begin{equation}
\kappa_n \approx -  \frac{\cos\Theta}{R_0}[1 +  \mathcal{O}(\frac{1}{A})] - \sum_i
M_i \frac{\gamma_i}{R_0 r}  \cos(M_{i}\theta - N_{i}\zeta)  ,
\end{equation} 
\begin{equation}
\kappa_g \approx  \frac{\sin\Theta}{R_0}[1 +  \mathcal{O}(\frac{1}{A})] + \sum_i
M_i \frac{\gamma_i}{R_o r}  \sin(M_{i}\theta - N_{i}\zeta)  .
\end{equation}
Here, the first terms correspond to the usual $1/R$ contributions to the curvature
vector, while the last terms correspond to the weak modulations due to the 3-D fields.
While the 3-D fields provide very small corrections to the curvature vector, they
can produce substantial corrections to the Pfirsch-Schluter coefficient.   The
solution to Eq. (18) for this equilibrium is given by
\begin{equation}
\lambda \approx  2\mu_o\hat{V}' \frac{r}{R_0}[- q\cos\Theta -  \sum_i
\frac{M_i q}{M_i - N_i q} \frac{\gamma_i}{r} \cos(M_i\Theta - N_i\zeta)]  ,
\end{equation}
where the first term is the conventional high-aspect ratio prediction for the
Pfirsch-Schluter coefficient and the last term is due to the 3-D fields.  As noted,
the 3-D fields can produce an appreciable effect when the 3-D field is
near pitch resonance with the $q$ profile.   

With these modifications, the 3-D corrections to the shifted circle equilibrium 
can be included assuming $\gamma_i/r \ll 1$ but $(\gamma_i/r)(M_i - N_i q)^{-1}
\sim \mathcal{O}(1)$.   The ballooning equation for the shifted-circle
equilibrium with small 3-D distortions is given by
\begin{equation}
\frac{\partial}{\partial\Theta}[(1 + \Lambda^2)\frac{\partial\xi}{\partial\Theta}]
+ \alpha[\cos\Theta + \Lambda \sin\Theta] \xi  =  - (1 + \Lambda^2) \hat{\omega}^2 \xi ,
\end{equation}
where the effect of the 3-D fields enters through the integrated local shear
quantity $\Lambda$ given by
\begin{equation}
\Lambda = \int_{\Theta_k}^{\Theta}  d\Theta \{s -  \alpha(\cos\Theta
+ \sum_i  \frac{M_i}{M_i - N_iq} \frac{\gamma_i}{r} \cos[(M_i - N_i q)\Theta - N_i \alpha_0)\} ,
\end{equation}
where $\alpha_0$ denotes a field line label.  Generally, numerical solutions
show that the large variation of the local shear produced by the 3-D fields
produces localized eigenfunctions along the field line.  Using this strong ballooning
approximation, the ballooning equation can be converted to an equation
of the form $d^2Y/d\Theta^2  + V(\Theta)Y =  - \hat{\omega}^2Y$, where the potential function $V$ can be expanded about $\Theta = 0$, $V
=  V(0)  +  V''(0) \Theta^2/2 + ...$.   From this calculation, the corresponding critical $\alpha$ for ballooning
instability can be calculated.  In the large $\alpha$, $s$ limit, this is given approximately by
\begin{equation}
\alpha_{crit}  \approx  \frac{s}{1 + \sum_i \frac{M_i}{M_i - N_i q} \frac{\gamma_i}{r} \cos(N_i\alpha_0)} \pm .... ,
\end{equation}
where higher order corrections are suppressed.  From this, we can see that the
 3-D fields can produced order unity corrections to ballooning stability boundary when
 $\gamma_i/r \sim (M_i - N_iq)$.   When this is the case, the $\alpha_{crit}$ can be 
 substantially modified.  Additionally, we note that the 3-D correction is
 sensitive to field-line label as noted by the $\alpha_0$ dependence.   Both of these
 features are also demonstrated by the more detailed numerical study described in the
 bulk of the paper.

\begin{center}
\bf{ACKNOWLEDGMENTS}
\end{center}

This research was supported by the U.S. Department of Energy under grant
nos.\ DE-FG02-99ER54546 and DE-FG02-86ER53218.   The authors would 
like to thank
J. D. Callen for useful discussions.


\begin{thebibliography}{11}

\bibitem{boozer}
A. H. Boozer, Phys. Plamsas \textbf{16}, 058102 (2009).

\bibitem{callen}
J. D. Callen, Nucl. Fusion \textbf{51}, 094026 (2011).

\bibitem{evans1}
T. E. Evans et al., Phys. Rev. Letters \textbf{92}, 235003 (2004).

\bibitem{evans2}
T. E. Evans et al., Nature Physics \textbf{2}, 419 (2006).

\bibitem{evans3}
T. E. Evans et al., Nucl. Fusion \textbf{48}, 024002 (2008).

\bibitem{loarte}
A. Loarte et al., Plasma Phys. Controlled Fusion \textbf{45}, 1549 (2003).

\bibitem{snyder}
P. B. Snyder et al., Phys. Plasmas \textbf{9}, 2037 (2002).

\bibitem{wilson}
H. R. Wilson et al., Phys. Plasmas \textbf{9}, 1277 (2002).

\bibitem{fitzpatrick}
R. Fitzpatrick, Nucl. Fusion \textbf{33}, 1049 (1993)

\bibitem{marsf_shielding}
Y. Q. Liu et al., Nucl. Fusion \textbf{51}, 083002 (2011).

\bibitem{moyers}
R. A. Moyer et al., Phys. Plasmas \textbf{12}, 056119 (2005).

\bibitem{Liang}
Y. Liang et al., Nuc. Fusion \textbf{50}, 025013 (2010).

\bibitem{denner}
P. Denner et al., Nucl. Fusion \textbf{52}, 054007 (2012).

\bibitem{suttrop}
W. Suttrop et al., Phys. Rev Letters \textbf{106}, 225004 (2011).

\bibitem{pueschel}
M. J. Pueschel et al., Phys. Plasmas {\bf 15}, 102310 (2008).

\bibitem{kbm}
P. B. Snyder and G. W. Hammett, Phys. Plasmas \textbf{8}, 744 (2001).

\bibitem{belli}
E. A. Belli and J. Candy, Phys. Plasmas \textbf{17}, 112314 (2010).

\bibitem{EPED}
P. B. Snyder et. al., Phys. Plasmas \textbf{16}, 056118 (2009).

\bibitem{groebner}
R. J.  Groebner et al., Nucl. Fusion \textbf{49}, 085037 (2009).

\bibitem{dickinson1}
D. Dickinson et al., Plasma Phys. Controlled Fusion \textbf{53} 115010 (2011).

\bibitem{dickinson2}
D. Dickinson et al., Phys. Rev. Letters \textbf{108}, 135002 (2012).

\bibitem{snyder2}
P. B.  Snyder et al., Phys. Plasmas \textbf{19}, 056115 (2012).

\bibitem{local3D}
C. C. Hegna, Phys. Plasmas \textbf{7}, 3921 (1999).

\bibitem{hh}
S. R. Hudson and C. C. Hegna, Phys. Plasmas \textbf{10}, 4716 (2003).

\bibitem{nakajima06}
N. Nakajima et. al., Nucl. Fusion \textbf{46}, 177 (2006).  

\bibitem{miller}
R. L. Miller et al., Phys. Plasmas \textbf{5}, 973 (1998).

\bibitem{3dballoon}
C. C. Hegna and N. Nakajima, Phys. Plasmas \textbf{5}, 1336 (1998).

\bibitem{snyder3}
P. B. Snyder et al., Phys. Plasmas \textbf{16}, 056118 (2009).

\bibitem{cht}
J. W. Connor, R. J. Hastie and J. B. Taylor, Phys. Rev. Lett. {\bf 40}, 396 (1978).

\end{thebibliography}
\end{document}